\documentstyle[12pt]{article}
\newcommand{\be}{\begin{equation}}
\newcommand{\ee}{\end{equation}}
\newcommand{\bea}{\begin{eqnarray}}
\newcommand{\eea}{\end{eqnarray}}
\newcommand{\nn}{\nonumber}

\input epsf
\begin{document}
\def\slash#1{\setbox0=\hbox{$#1$}#1\hskip-\wd0\dimen0=5pt\advance
       \dimen0 by-\ht0\advance\dimen0 by\dp0\lower0.5\dimen0\hbox
         to\wd0{\hss\sl/\/\hss}}\def\ink {\int~{d^4k\over (2\pi)^4}~}

\def\gamh{\Gamma_H}
\def\esp #1{e^{\displaystyle{#1}}}
\def\de{\partial}
\def\eb{E_{\rm beam}}
\def\deb{\Delta E_{\rm beam}}
\def\sigm{\sigma_M}
\def\sigmmax{\sigma_M^{\rm max}}
\def\sigmmin{\sigma_M^{\rm min}}
\def\sige{\sigma_E}
\def\dsigm{\Delta\sigma_M}
\def\mh{M_H}
\def\lyear{L_{\rm year}}

\def\wstar{W^\star}
\def\zstar{Z^\star}
\def\ie{{\it i.e.}}
\def\etal{{\it et al.}}
\def\eg{{\it e.g.}}
\def\pzero{P^0}
\def\mt{m_t}
\def\mpzero{M_{\pzero}}
\def\mev{~{\rm MeV}}
\def\gev{~{\rm GeV}}
\def\gam{\gamma}
\def\lsim{\mathrel{\raise.3ex\hbox{$<$\kern-.75em\lower1ex\hbox{$\sim$}}}}
\def\gsim{\mathrel{\raise.3ex\hbox{$>$\kern-.75em\lower1ex\hbox{$\sim$}}}}
\def\ntc{N_{TC}}
\def\epem{e^+e^-}
\def\tauptaum{\tau^+\tau^-}
\def\lplm{\ell^+\ell^-}
\def\anti{\overline}
\def\mw{M_W}
\def\mz{M_Z}
\def\fbi{~{\rm fb}^{-1}}
\def\mupmum{\mu^+\mu^-}
\def\rts{\sqrt s}
\def\sigrts{\sigma_{\tiny\rts}^{}}
\def\sigrtssq{\sigma_{\tiny\rts}^2}
\def\sigrtsprime{\sigma_{E}}
\def\nsigrts{n_{\sigrts}}
\def\gampzero{\Gamma_{\pzero}}
\def\pzerop{P^{0\,\prime}}
\def\mpzerop{M_{\pzerop}}

\font\fortssbx=cmssbx10 scaled \magstep2
%
%
%
\hfill
$\vcenter{
 \hbox{\bf BARI-TH 398/00}
\hbox{\bf DFF-365/10/00} \hbox{\bf UGVA-DPT-2000/09-1089} }$
\medskip
\begin{center}
{\Large\bf\boldmath {Dispersion Laws for In-medium Fermions
}}\vskip0.3cm {\Large\bf\boldmath{ and Gluons in the CFL Phase of
QCD }}\\ \rm \vskip1pc {\Large R. Casalbuoni$^{a,b}$,  R.
Gatto$^c$ and G. Nardulli$^{d,e}$}\\ \vspace{5mm}
{\it{$^a$Dipartimento di Fisica, Universit\`a di Firenze, I-50125
Firenze, Italia
\\
$^b$I.N.F.N., Sezione di Firenze, I-50125 Firenze, Italia\\
$^c$D\'epart. de Physique Th\'eorique, Universit\'e de Gen\`eve,
CH-1211 Gen\`eve 4, Suisse\\ $^d$Dipartimento di Fisica,
Universit\`a di Bari, I-70124 Bari, Italia\\ $^e$I.N.F.N., Sezione
di Bari, I-70124 Bari, Italia }}
\end{center}
\bigskip
\begin{abstract}

We evaluate several quantities appearing in the effective
lagrangian for the color-flavor locked phase of high density QCD
using a formalism which exploits the approximate decoupling of
fermions with energy negative with respect to the Fermi energy.
The effective theory is essentially two-dimensional and exhibits a
Fermi velocity superselection rule, similar to the one found in
the  Heavy Quark Effective Theory. Within the formalism we
reproduce, using gradient expansion, the results for the effective
parameters of the Nambu-Goldstone bosons. We also determine the
dispersion laws for the gluons. By coupling the theory to fermions
and integrating over the two-dimensional degrees of freedom we
obtain the effective description of in-medium fermions.

\noindent

\end{abstract}

\section{Introduction}

QCD at high density is expected to be in a chirality breaking
color superconducting phase due to formation of quark Cooper pairs
\cite{a},\cite{alford}. In QCD with three massless quarks one
expects simultaneous breaking of color and flavor simmetries into
a residual diagonal color-flavor locked symmetry (CFL). The
physical description and consequences have been discussed by a
number of authors \cite{b,casalbuoni,rho,son}. A formalism has
been recently proposed \cite{hong} and \cite{beane} to deal with
the color superconducting phase. It takes advantage of the
approximate decoupling of the states of a massless fermion of
energy distant
 with respect to the Fermi surface. We shall first summarize the
main consequences of such a description, with emphasis on its
implied velocity superselection rule, similar to that familiar in
the Heavy Quark Effective Theory (HQET) \cite{HQET}, and on its
essentially two-dimensional features, due to the special relevance
of momenta parallel to the Fermi velocity. We shall then calculate
in this formalism the effective parameters of the goldstones,
reobtaining results derived by other methods. Next we shall
calculate the dispersion laws for the gluons and discuss the
various masses that can be defined for the scalar, longitudinal
and transverse gluons, i.e. their rest mass,  the effective mass
and the inverse of the penetration length. After having coupled
the theory to fermions we obtain the effective description of
in-medium fermions, by integrating over the two-dimensional
degrees of freedom.

 In section 2 we discuss the effective theory
on the Fermi surface. In section 3 we write down the effective
lagrangian for the fermions and the goldstones. In section 4 we
perform the gradient expansion on the generating functional. We
discuss the gluon polarization tensor in section 5. In section 6
we treat the effective description of in-medium fermions.

\section{Effective theory on the Fermi surface}

As discussed in the introduction, we will make use of the
formalism developed in refs. \cite{hong,beane} to evaluate several
quantities appearing in the effective lagrangian \cite{casalbuoni}
describing the CFL phase of QCD at high density. This formulation
is based on the observation that, at very high-density, the energy
spectrum of a massless fermion is described by states
$|\pm\rangle$ with energies \be E_\pm =-\mu\pm|\vec p\,|~,\ee
where $\mu$ is the is the quark number chemical potential. For
energies much lower than the Fermi energy $\mu$, only the states
$|+\rangle$ close to the Fermi surface. i.e. with  $|\vec
p\,|\approx\mu$, can be excited. On the contrary, the states
$|-\rangle$ have $E_-\approx -2\mu$ and therefore they decouple
leaving in the physical spectrum only the states $|+\rangle$ and
the gluons. This can be seen more formally by writing the
four-momentum of the fermion as \be p^\mu=\mu
v^\mu+\ell^\mu\label{momentum}~,\ee where $v^\mu=(0,\vec v_F)$,
with $|\vec v_F|=1$. Since the hamiltonian for a massless Dirac
fermion in a chemical potential $\mu$ is \be
H=-\mu+\vec\alpha\cdot\vec
p\,,~~~\vec\alpha=\gamma_0\vec\gamma~,\ee one has \be
H=-\mu(1-\vec\alpha\cdot \vec v_F)+\vec \alpha\cdot\vec\ell~.\ee
Then, it is convenient to introduce the projection operators \be
P_\pm=\frac{1\pm\vec\alpha\cdot\vec v_F}2~,\ee such that \be
H|+\rangle=\vec\alpha\cdot\vec\ell\, |+\rangle,~~~
H|-\rangle=(-2\mu+\vec\alpha\cdot\vec\ell )\,|-\rangle~. \ee
 In
terms of fields, the momentum decomposition in eq.
(\ref{momentum}) and the projection of the states $|\pm\rangle$
can be realized through \be \psi(x)=\sum_{\vec v_F}\,\esp{-i\mu
v\cdot x}\left[\psi_+(x)+\psi_-(x)\right]~,\ee where
$\displaystyle \sum_{\vec v_F}$ is an average over the Fermi
velocities, and the velocity-dependent fields $\psi_\pm (x)$ are
given by:
 \be \psi_\pm(x)=\esp{i\mu v\cdot
x}\left(\frac{1\pm\vec\alpha\cdot\vec
v_F}2\right)\psi(x)~=~\int_{|\ell
|<\mu}\frac{d^4\ell}{(2\pi)^4}\esp{-i\,\ell\cdot
x}\psi_\pm(\ell)~. \label{psimeno}\ee Introducing the four-vectors
\be V^\mu=(1,\vec v_F),~~~\tilde V^\mu=(1,-\vec v_F)~, \ee it is
easy to derive the following identities \bea
\bar\psi_+\gamma^\mu\psi_+&=&\psi_+^\dagger V^\mu\psi_+\nn\\
\bar\psi_-\gamma^\mu\psi_-&=&\psi_-^\dagger {\tilde
V}^\mu\psi_-\nn\\ \bar\psi_+\gamma^\mu\psi_-&=&\bar\psi_+
\gamma^\mu_\perp\psi_-\nn\\
\bar\psi_-\gamma^\mu\psi_+&=&\bar\psi_- \gamma^\mu_\perp\psi_+~ .
\label{iden} \eea
where
\be
\gamma^\mu_\perp~=~\frac 1 2 \gamma_\nu\left(2 g^{\mu\nu}-
V^\mu\tilde V^\nu-\tilde V^\mu V^\nu \right)
 ~. \ee
Substituting into the Dirac part of the QCD
lagrangian density one obtains
\be
 {\cal L}=\sum_{\vec v_F}
\left[\psi_+^\dagger iV\cdot D\psi_++\psi_-^\dagger(2\mu+ i\tilde
V\cdot D)\psi_-+(\bar\psi_+i\slash D_\perp\psi_- + {\rm h.c.}
)\right]~,\label{expression} \ee where $\slash
D_\perp=D_\mu\gamma^\mu_\perp$
 and $D_\mu$ is the usual covariant derivative.
 We notice that the fields appearing in (\ref{expression})
  are evaluated at the same Fermi velocity because
off-diagonal terms are cancelled by the rapid oscillations of the
exponential factor in the $\mu\to\infty$ limit. This behaviour is similar
to that in the Heavy
Quark Effective Theory \cite{HQET}, and can be referred to, by
analogy, as the Fermi velocity superselection rule.

At the leading order in $1/\mu$ one has
 \be iV\cdot
D\, \psi_+=0,~~~~~~\psi_-~=~-i~\frac{1}{2\mu}\gamma_0 \, \slash
D_\perp\, \psi_+~,\label{eq12}\ee showing the decoupling of
$\psi_-$ in the $\mu\to\infty$ limit. Therefore, in this limit,
the $\psi_-$ field plays no role, except in those loop diagrams,
of the order of $\mu^3$ instead of $\mu^2$, where the $1/\mu$
factor in (\ref{eq12}) is eaten by the  extra $\mu$ factor from
the momentum integration. We shall see an example in Section 5.
 The equation for $\psi_+$ shows
also that only the energy and the momentum parallel to the Fermi
velocity are relevant variables in the problem. We have an
effective two-dimensional theory.

In conclusion at the next to leading order in $\mu$ we have
\be
 {\cal L}=\sum_{\vec v_F}
\left[\psi_+^\dagger iV\cdot D\psi_+  -
\frac{1}{2\mu}\psi_+^\dagger( \slash D_\perp)^2\psi_+\right]~.
\label{eff} \ee Further terms in the $1/\mu$ expansion can be
found in \cite{hong}.

 The previous remarks apply to any theory
describing massless fermions at high density. In the case of CFL
phase of QCD, in order to implement correctly the symmetry
properties, one has to consider left- and right-handed fermions
transforming respectively as the ${\bf{(3,1,3)}}$ and
${\bf{(1,3,3)}}$ representations of $SU(3)_L\otimes SU(3)_R\otimes
SU(3)_c$. Therefore we will consider for each left- and
right-handed field an effective two-dimensional theory as outlined
above. The next step will be to couple this theory in a
$SU(3)_L\otimes SU(3)_R\otimes SU(3)_c$ invariant way to
Nambu-Goldstone bosons (NGB)  describing the appropriate breaking
\be SU(3)_L\otimes SU(3)_R\otimes SU(3)_c\to SU(3) \ee for the CFL
phase. Using a gradient expansion we will get an explicit
expression for  the decay coupling constant of the Nambu-Goldstone
boson as well as for its velocity, whose value $v=1/\sqrt 3$ is
interpreted as a consequence of the average over the Fermi
velocities. These quantities have been obtained previously by
different methods \cite{son}. By the same formalism we will
determine the dispersion law for the gluons in the $\mu\to\infty$
limit. Finally we will couple the theory to external fermions.
Integrating out the two-dimensional degrees of freedom we will get
an effective theory describing NGB's  and the in-medium fermions.

\section{Couplings to the Goldstone fields}

Let us recall the effective description of the CFL phase given in
ref. \cite{casalbuoni}.  The symmetry breaking is induced by the
condensates \be \langle\psi_{ai}^L\psi_{bj}^L\rangle=
-\langle\psi_{ai}^R\psi_{bj}^R\rangle=\gamma_1\,\delta_{ai}\delta_{bj}+
\gamma_2\,\delta_{aj} \delta_{bi}~,\label{condensates}\ee where
$\psi_{ai}^{L(R)}$ are Weyl spinors and a sum over spinor indices
is understood. The indices $a,b$ and $i,j$ refer to $SU(3)_c$ and
$SU(3)_L$ (or $SU(3)_R$) respectively. One then  introduces
$SU(3)$ matrix-valued fields $X$ and $Y$ transforming under the
symmetry group $SU(3)_c\otimes SU(3)_L\otimes SU(3)_R$ as
left-handed and right-handed fermions respectively. That is \be
X\to g_c Xg_L^T,~~~~Y\to g_c Y g_R^T~.\ee Using the gauge freedom
associated to $SU(3)_c$ it is possible to choose a gauge such that
$X=Y^\dagger$. The condensates of eq. (\ref{condensates}) break
also $U(1)_V\otimes U(1)_A$ and as a consequence one also needs to
introduce the related Goldstone fields. However we will not
discuss these fields in this paper since all the calculations
presented here are trivially extended to them \cite{son}.

The invariant coupling between fermions and Goldstone fields
reproducing the symmetry breaking pattern of eq.
(\ref{condensates}) is proportional to \be
\gamma_1\,Tr[\psi_L^TX^\dagger]\,C\, Tr[\psi_L
X^\dagger]+\gamma_2\,Tr[\psi_L^TCX^\dagger\psi_L X^\dagger]+{\rm
h.c.}~,\label{invariant}\ee and analogous relations for the
right-handed fields. Here the spinors are meant to be Dirac
spinors and $C=i\gamma^2\gamma^0$ is the charge-conjugation
matrix. The trace is operating over the group indices of the
spinors and of the Goldstone fields. Since the vacuum expectation
value of the Goldstone fields is $\langle X\rangle=\langle
Y\rangle=1$, we see that this coupling induces the correct
breaking of the symmetry. In the following we will consider only
the case $\gamma_2=-\gamma_1\propto\Delta/2$, where $\Delta$ is
the gap parameter. This choice corresponds to a condensate
behaviour as the $\bf{(3,3)}$  representation of $SU(3)_c\otimes
SU(3)_{L,R}$. This situations seems the one which is favorite by
the dynamical analysis made in ref. \cite{alford}. Again, the
calculations presented here can be easily extended to the more
general situation where the condensate behaves as $\bf{(3,3)\oplus
(6,6)}$. By our choice here the condensate has the form \be
\langle\psi_{ai}^L\psi_{bj}^L\rangle=
-\langle\psi_{ai}^R\psi_{bj}^R\rangle={\gamma_1}\,
\epsilon_{abI}\epsilon_{ijI}~,\ee and the invariant coupling of
eq. (\ref{invariant}) can be written as \be -\frac{\Delta} 2
\sum_{I=1,3}Tr[(\psi X^\dagger)^T C\epsilon_I(\psi
X^\dagger)\epsilon_I]~, \ee where, now, also $\psi$ is considered
as a $3\times 3$ matrix and the matrix $\epsilon_I$ is defined as
\be (\epsilon_I)_{ab}=\epsilon_{Iab}~. \ee Notice that for any
$SU(3)$ matrix $g$ one has the identity \be g^T\epsilon_I
g=\epsilon_{I'}(g^\dagger)_{I'I}~. \ee Then it is easy to show
that our expression for the coupling is equivalent to the one used
by the authors of ref. \cite{rho} \be
-\frac{\Delta}2\sum_{I=1,3}Tr[(\psi X^\dagger)^T C\epsilon_I(\psi
X^\dagger)\epsilon_I]=-\frac{\Delta}2\sum_{I,I'=1,3}
Tr[\psi^TC\epsilon_I X_{II'}\psi\epsilon_{I'}]~. \ee

Since the transformation properties under the symmetry group of
the fields at fixed Fermi velocity, as introduced in the previous
section, do not differ from those of the  quark fields, for both
left-handed and right-handed fields we get the effective
lagrangian density \bea {\cal L}_1&=&\sum_{\vec v_F} \frac 1 2
\Big[\sum_{A=1}^9\left(\psi_+^{A\dagger}iV\cdot
D\psi_+^A+\psi_-^{A\dagger}i\tilde V\cdot
D\psi_-^A-{\Delta_A}\left({\psi_-^A}^TC\psi_+^A+{\rm
h.c.}\right)\right)\cr&-&{\Delta}\sum_{I=1,3}\left(Tr[(\psi_-
X_1^\dagger)^T C\epsilon_I(\psi_+
X_1^\dagger)\epsilon_I]+\rm{h.c.}\right)\Big]\label{lagrangian}~,
\eea where we have introduced the fields
\be\psi_\pm=\frac{1}{\sqrt{2}}\sum_{A=1}^9\lambda_A\psi^A_\pm~,\ee
and $\lambda_a~(a=1,...,8)$ are the Gell-Mann matrices normalized
as follows: $Tr(\lambda_a\lambda_b)=2\delta_{ab}$ and
$\lambda_9=\sqrt{ 2/ 3}~{\bf {1}}$. Furthermore
$\Delta_1=\cdots=\Delta_8=\Delta$ and $\Delta_9=-2\Delta$ and
$X_1=X-1$. Notice that  the NGB fields couple to fermionic fields
with opposite Fermi velocities. In this expression, as in the
following ones, the field $\psi_-$ is defined as $\psi_+$ with
$\vec v_F\to-\vec v_F$, and therefore it is not the same as the
one defined in (\ref{psimeno}). In the previous lagrangian we have
also separated the free quadratic terms in the fermionic fields.
These terms can be rewritten in a more compact way by introducing
the following fields \be \chi=\left(\matrix{\psi_+\cr
C\psi^*_-}\right)~,\ee It is important to realize that the fields
$\chi$ and $\chi^\dagger$ are not independent variables. In fact,
since we integrate over all the Fermi surface, the fields
$\psi_-^*$ and $\psi_+$, appearing in $\chi$, appear also in
$\chi^\dagger$ when $\vec v_F\to -\vec v_F$. In order to avoid
this problem we can integrate over half of the Fermi surface, or,
taking into account the invariance under $\vec v_F\to -\vec v_F$,
we can simply integrate over all the sphere with a weight $1/8\pi$
instead of $1/4\pi$: \be \sum_{\vec v_F}=\int\frac{d\vec
v_F}{8\pi}~. \ee Then the first three terms in the lagrangian
density (\ref{lagrangian}) become \be {\cal L}_0=\int\frac{d\vec
v_F}{8\pi}~ \frac 1 2\sum_{A=1}^9
\chi^{A\dagger}\left[\matrix{iV\cdot D & \Delta^A\cr\Delta^A
&i\tilde V\cdot D^*}\right]\chi^A\label{kinetic}~, \ee so that, in
momentum space the free  fermion propagator is \be
S_{AB}(p)=\frac{2\delta_{AB}}{V\cdot p\,\tilde V\cdot
p-\Delta_A^2}\left[\matrix{\tilde V\cdot p & -\Delta_A\cr-\Delta_A
& V\cdot p}\right]~.\ee  We note explicitly that at this stage,
since we have integrated out the degrees of freedom of the gluons
in order to obtain the condensate appearing in eq.
(\ref{lagrangian}), the gluon field appearing in the covariant
derivative in (\ref{kinetic}) has to be interpreted as an
external field.  From eq. (\ref{kinetic}) we can read also the
fermionic current which is given by \bea J_\mu^a&=&\frac 1 4\Big(
\sqrt{\frac 2 3}
\sum_{a=1}^8\left(\chi^{9\dagger}\left[\matrix{V_\mu & 0\cr 0
&-\tilde V_\mu}\right]\chi^a+{\rm h.c.}\right)\cr &+&
\sum_{b,c=1}^8\chi^{b\dagger} \left[\matrix{V_\mu g_{bac} & 0\cr 0
&-\tilde V_\mu g_{bac}^*}\right]\chi^c\Big)~, \label{current} \eea
where \be g_{abc}=d_{abc}+if_{abc}~,\ee and the symbols
$d_{abc}~,f_{abc}$ are defined, as usual, by $\displaystyle
\lambda_a\lambda_b+\lambda_b\lambda_a =2d_{abc}\lambda_c+\frac 4 3
\delta_{ab}$ and $\displaystyle
\lambda_a\lambda_b-\lambda_b\lambda_a=2if_{abc}\lambda_c~$.

Moreover, expanding the NGB fields \be X=
 \exp{\displaystyle{ i\left(
\frac{\lambda_a\Pi^a}{2F}\right)}},~~~~~~a=1,\cdots,8~,\ee we get,
considering terms up to the second order in the Goldstone fields:
 \be X_1=\, i
\frac {\lambda_a \Pi^a}{2F}\,-  \frac {\lambda_a \lambda_b \Pi^a
\Pi^b} {8F^2}~. \ee Using these expressions, the trilinear
coupling of the NGB's to the fermions arising from the last term
in (\ref{lagrangian})  can be written as

\bea &&{\cal L}_{\chi\chi\Pi} = \int \frac{d\vec v_F}{8\pi}
\frac{i\,\Delta}{2 F} \times \cr &&  \left(\sum_{a=1}^8
\left(\chi^{9\dagger}\left[\matrix{0 & -1\cr 1
&0}\right]\chi^a\frac{\Pi^a}{\sqrt 6} + {\rm h.c.} \right) -
\sum_{a,b,c=1}^8 d_{abc}\chi^{a\dagger} \left[\matrix{0 & -1\cr 1
&0}\right]\chi^b\Pi^c\right) ~.\eea

The quadrilinear coupling of two Goldstone bosons to two fermions
is \bea
 &&{\cal L}_{\chi\chi\Pi\Pi}=\int \frac{d\vec v_F}{8\pi}
 \Big\{~\frac{4}{3} \sum_{a=1}^8 \frac{\Delta}{16F^2}
\chi^{9\dagger} \left[\matrix{0 & +1\cr 1 &0}\right]
\chi^9~\Pi^a\Pi^a~ \nonumber
\\&& + 3~\sqrt{\frac{2}{3}}
\sum_{a,b,c=1}^8\left(\frac{\Delta}{16F^2} d_{abc}
\chi^{c\dagger}\left[\matrix{0 & +1\cr 1
&0}\right]\chi^9~\Pi^a\Pi^b + {\rm h.c.}\right)~\nonumber
\\ &&-~ \sum_{a,b,c,d=1}^8
\frac{\Delta}{16F^2}h_{abcd}\chi^{c\dagger} \left[\matrix{0 &
-1\cr 1 &0}\right] \chi^d\Pi^a\Pi^b
 \Big\}~,\eea
where \be h_{abcd}=2 \sum_{p=1}^8
\left(g_{cap}g_{dbp}+d_{cdp}d_{abp}\right) \, -\frac 8 3
\delta_{ac}\delta_{db} +\frac 4 3 \delta_{cd}\delta_{ab}~.\ee

\section{Gradient expansion}

We will treat the Goldstone bosons as external fields and we will
perform a derivative expansion of the generating functional. This
will give rise to the effective action for the NGB's. At the
lowest order the relevant diagrams with two external NGB lines are
the ones in Fig. 1.
\begin{figure}[htb]
\epsfxsize=8truecm \centerline{\epsffile{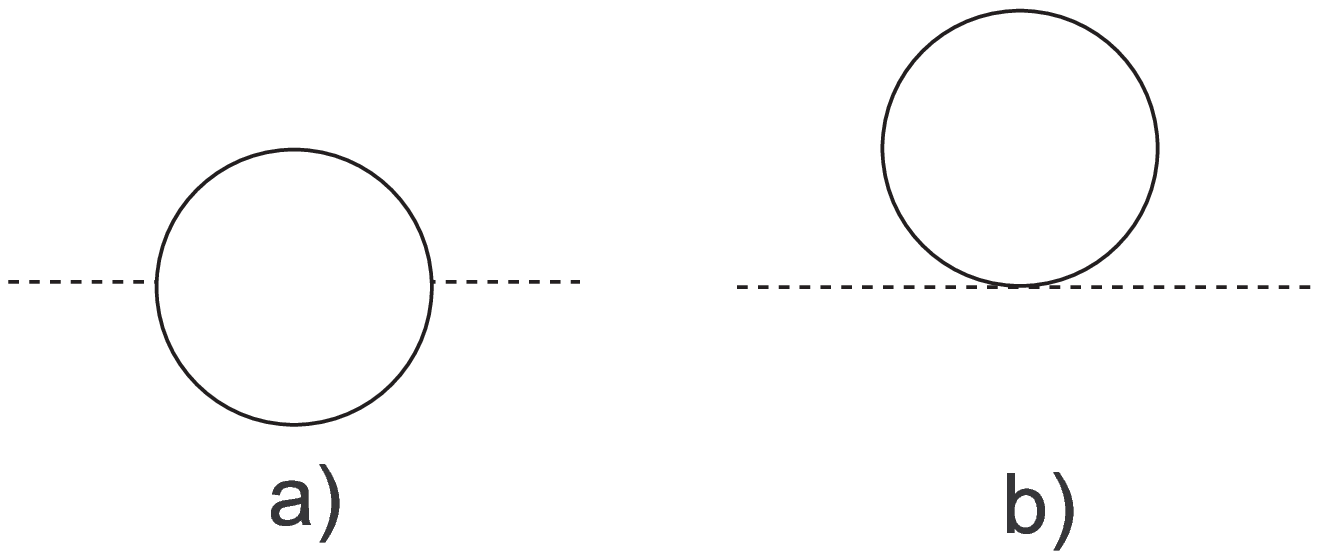}} \noindent
{\bf Fig. 1} - {One-loop diagrams in the gradient expansion.
Dotted lines are the external fields, full lines are fermion
propagators.}
\end{figure}
As previously discussed, since the fermions live effectively near
the Fermi surface, the four-dimensional momentum integration can
be reduced to a two-dimensional one according to the formula \be
\int\frac{d^4\ell}{(2\pi)^4}=\frac{\mu^2}{\pi}
\int\frac{d^2\ell}{(2\pi)^2}~, \ee where $d^2\ell=d\ell_0
d\ell_{\|}$; notice that effectively the integration limits are
$(-\mu,+\mu)$.
 In the Feynman rules one has also to take into
account a factor 2 coming from the contribution of left-handed and
right-handed fermions.

The tadpole diagram contributes only to the mass term and it is
essential to cancel the external momentum independent  term
arising from the other diagram. Therefore, as expected, the mass
of the NGB's is zero. The contribution at the second order in the
momentum expansion is given by
 \be
i\,  \frac{21-8\ln 2}{72\pi^2 F^2} \int\frac{d\vec
v_F}{4\pi}\sum_{a=1}^8 \Pi^a\,V\cdot p\, \tilde V\cdot p\,\Pi^a~.
\ee Integrating over the velocities and going back to the
coordinate space we get \be{\cal L}_{\rm eff}^{\rm kin}=
\frac{21-8\ln 2} {72\pi^2 F^2}\sum_{a=1}^8
\left(\dot\Pi^a\dot\Pi^a-\frac 1 3|\vec\nabla\Pi_a|^2\right)\, .
\ee We can now determine the decay coupling constant $F$ through
the requirement of getting the canonical normalization for the
kinetic term; this implies, {\it non nova sed nove},\be F^2=\frac
{\mu^2(21-8\ln
2)}{36\pi^2}\, ,\ee a result already obtained by other authors
with a different method.

 In this formalism the origin of the goldstone
velocity $1/\sqrt{3}$ is a direct consequence of the integration
over the Fermi velocity. Therefore it is completely general and
applies to all the NGB's in the theory, including the ones
associated to the breaking of $U(1)_V$ and $U(1)_A$; needless to
say, higher order terms in the expansion $1/\mu$ could change
this result.

The breaking of the Lorentz invariance, exhibited by the pion
velocity being different from one, can be seen also in the matrix
element $\langle 0|J_\mu^a|\Pi^b\rangle$. This can be evaluated
using the Feynman diagram of Fig. 2.
\begin{figure}[htb]
\epsfxsize=4truecm \centerline{\epsffile{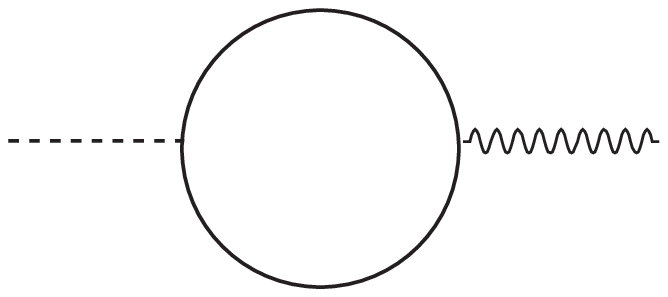}} \noindent
{\bf Fig. 2} - {Diagram contributing to the matrix element of the
fermionic current between the CFL vacuum  and the single NGB
state. The  dotted line represents the Goldstone boson, the wavy
line the current and the internal lines correspond to fermions.}
\end{figure}
The result, after the integration over the Fermi velocity is
\be\langle 0|J_\mu^a|\Pi^b\rangle=iF\delta_{ab} \tilde p_\mu,
~~~~\tilde p^\mu=(p^0,~\frac 1 3 \, \vec p)~,\ee where we have
made use of the previous result for $F$. It is interesting to
notice that the current is conserved,  as the dispersion law for
the NGB's  has the form $\displaystyle{p\cdot\tilde
p=p_0^2-1/3{|\vec p}\,|^2=0}$.

\section{Dispersion law for the gluons}

The dispersion law for the in-medium gluons can be derived by the
same formalism used in the previous section. It is equivalent to
consider the polarization tensor $\Pi^{\mu\nu}(p)$. Its one-loop
contribution is obtained by computing diagrams analogous to the
ones in Fig. 1 (with external lines representing currents
$J_\mu^a,~J_\nu^b$ as given by (\ref{current})). The result of
the first diagram (Fig.1 a) is:
\begin{equation}
\Pi^{\mu\nu}_{ab}(p)\Big
|_{1a}=\Pi^{\mu\nu}_{ab}(0)+\delta\Pi^{\mu\nu}_{ab}(p)
\label{pimunu}
\end{equation} where \be \Pi^{\mu\nu}_{ab}(0)=\frac{\mu^2
g_s^2}{4\pi}\delta_{ab} \int\frac{d\vec
v_F}{4\pi}\Sigma^{0,\mu\nu}~,\ee
 \be
\delta\Pi^{\mu\nu}_{ab}(p)=\frac{\mu^2 g_s^2}{4\pi}\delta_{ab}
\int\frac{d\vec v_F}{4\pi}\Sigma^{\mu\nu}~,\ee and
\be\Sigma^{0,\mu\nu}=k_1(V^\mu V^\nu+\tilde V^\mu\tilde
V^\nu)+k_2(V^\mu\tilde V^\nu+\tilde V^\mu V^\nu)~,\ee
\be\Sigma^{\mu\nu}=a\left(V^\mu V^\nu\frac{(\tilde V\cdot
p)^2}{\Delta^2}+\tilde V^\mu\tilde V^\nu\frac{( V\cdot
p)^2}{\Delta^2}\right)+b\left(V^\mu\tilde V^\nu+\tilde V^\mu
V^\nu\right)\frac{V\cdot p\,\tilde V\cdot p}{\Delta^2}\ .\ee In the
previous formulas we have included terms up to the second order in
momenta. The coefficients are given by \bea a&=&\frac{1}{108\pi}
\left(31-\frac{32}{3}\ln 2 \right)
\\
b&=&\frac{10 }{108\pi} \left(1-\frac{8}{3}\ln 2 \right)
\\
k_1&=&\frac{3 }{2\pi}
\\ k_2&=&-\,  \frac{1}{9\pi}
\left(3+4\ln 2 \right)~.
 \eea
To this result one should add the contribution arising from the
tadpole diagram of Fig. 1b (with, also in this case,  external
lines representing fermionic  currents $J_\mu^a,~J_\nu^b$). This
diagram arises from the last term in (\ref{eff}). It can be
computed by performing first the integration over the energy
$\ell_0$ and subsequently over the longitudinal momentum
$\ell_{\|}$ \cite{son}. This integration introduces an extra
factor $\mu$ besides the $\mu^2$ factor arising from the
integration over the Fermi surface. The result is independent on
the external momentum $p$ and is given by:  \be
\Pi^{\mu\nu}_{ab}(0)\Big
|_{1b}~=~\Delta\Pi^{\mu\nu}_{ab}(0)~=~\frac{\mu^2
g^2_s}{4\pi}\delta_{ab}\int\frac{d\vec
v_F}{4\pi}\Sigma^{0,\mu\nu}_B~, \ee with \be
\Sigma^{0,\mu\nu}_B=k_1\left[2 g^{\mu\nu}-V^\mu\tilde V^\nu
-V^\nu\tilde V^\mu \right]~.\ee From $\Pi^{\mu\nu}_{ab}(0)+\Delta
\Pi^{\mu\nu}_{ab}(0)$ we can read the one-loop contributions to
the Debye and Meissner mass \be m^2_D=g_s^2F^2=\frac{\mu^2
g_s^2}{36\pi^2} \left(21-8\ln 2 \right)~,\ee and \be
m^2_M=\frac{\mu^2 g_s^2}{\pi^2} \left(-\frac{11}{36}-\frac{2}{27}
\ln 2+\frac{1}{2} \right) \,=\,\frac{m^2_D}{3}~,
 \ee
where the
first two terms are the result of the diagram of Fig.1a and the
last one  is the result of the diagram of Fig. 1b (this
contribution is called the bare Meissner mass in \cite{son}).
These results agree with the findings of other authors \cite{son},
\cite{zarembo}. In particular $m_M^2 = v^2 m_D^2$, where $v$, the
NGB velocity, is equal to $1/\sqrt{3}$.

It may be noted, in passing, that the polarization tensor
$\Pi_{ab}^{\mu\nu}(p)\Big|_{1a+1b}$ satisfies a Ward identity in
the soft NGB limit. In fact, let us consider the amplitude $p_\nu
\Pi^{\mu\nu}_{ab}(p)$, with
$\Pi^{\mu\nu}_{ab}(p)=\Pi^{\mu\nu}_{ab}(p)\Big|_{1a+1b}$, in the
$p\to 0$ limit. In this kinematical regime it should be dominated
by the massless NGB pole and it should be proportional to
$\displaystyle{\langle 0|J_\mu^a|\Pi^b\rangle=iF\delta_{ab}\tilde
p_\mu}~$. In the $p\to 0$ limit we find indeed \be p_\nu
\Pi^{\mu\nu}_{ab}(p)\propto  \tilde
 p^\mu~. \ee For this behaviour to be
found it is essential to include both diagrams of Fig.1.

In order to derive the dispersion law for the gluons, we first
write the equations of motion for the gluon field $A^b_\mu$ in
momentum space and high-density limit: \be
\Pi^{\nu\mu}_{ab}\,A^b_\mu~=0~,
 \label{eqmot} \ee
from which we obtain \be p_\nu \Pi^{\nu\mu}_{ab}A^b_\mu~=0~, \ee
i.e. for all the gluons: \be p^0A_0=\frac{1}{3}(\vec p \cdot\vec
A)+...~,\label{eqmot2}
 \ee where
the ellipsis denotes  terms of the third order in the gluon
momenta that, consistently with our approximation, we neglect.
Substituting this result in ({\ref{eqmot}) taken for $\nu=0$ one
gets the dispersion law for the time-like gluons $A^b_0$ \be
3\alpha_1\,E^2-\alpha_1\,|\vec p\,|^2=m^2_D\ee with the factor 3
arising from the coupling of $A_0$ to $A_i$ in the equation of
motion and the use of eq. (\ref{eqmot2}). This gives \be p^0=\pm
\,E_{A_0},~~~E_{A_0}=\frac{ 1}{\sqrt 3}\sqrt{|\vec
p\,|^2+\frac{m^2_D}{\alpha_1}}~,\ee with \be \alpha_1=\frac{\mu^2
g_s^2}{6\Delta^2\pi}(a-b)=\frac{\mu^2
g_s^2}{216\Delta^2\pi^2}\left(7+\frac{16}{3}\ln 2\right)~.\ee
Notice that the rest mass of $A_0^b$ is given by  \be
m^R_{A_0}=\frac{m_D}{\sqrt{3\alpha_1}}= \sqrt{6\,\frac{21-8\ln
2}{21+16 \ln2}}\,\Delta\approx 1.70\,\Delta~.\ee This result may
appear surprising since the Debye and the Meissner masses are of
the order $g_s\mu$. However one should take into account that the
renormalization of the coefficient of $E^2$  (that is
$\alpha_1$), arising from the polarization tensor, redefines the
physical mass by a factor proportional to $\Delta/g_s\mu$  and
the coupling constant $g_s$ as well. Actually there are various
different "masses" which can be defined, apart from the rest
mass, when we consider the low momentum limit. For instance we
can consider the inverse of the penetration length, $m_{A_0}^P$,
defined for $E\to 0$ as the ratio between the mass term and the
coefficient of $|\vec p\,|^2$. This mass is given by \be
m^P_{A_0}=\sqrt{3}\,m^R_{A_0}\approx 2.94\,\Delta~.\ee We can
finally define a third mass, sometimes called effective mass,
$m^*$, by considering  the ratio between the spatial momentum and
the velocity, i.e. by looking at the following expression \be
\vec v=\frac{\partial E}{\partial\vec p}=\frac{\vec
p}{m^*(p)}~\ee in the limit  $\vec p\to 0$, that is \be
m^*=m^*(0)~.\ee The meaning of this last mass for ordinary
particles would be the coefficient appearing in the kinetic
energy term. One obtains\be m^*_{A_0}= \sqrt{\frac{3}{\alpha_1}}
m_D=3\,m^R_{A_0}\approx 5.10\, \Delta~. \ee

 Let us now turn to the dispersion law for $\vec A^a$; we
consider (\ref{eqmot}) for $\nu=i$ and use again (\ref{eqmot2}).
We introduce longitudinal and transverse components: \bea A_L^{i\,
a}&=&\frac{\vec p\cdot \vec A^a}{|\vec p|^2}\, p^i\cr A_T^{i\,
a}&=&A^{i\, a}-A_L^{i\, a}~,\eea and we project the equations of
motion along the longitudinal and transverse direction. The
results for the longitudinal and transverse cases are respectively
\bea \alpha_1\,E^2-\alpha_2\,\frac{|\vec
p\,|^2}3&=&\frac{m_D^2}3\cr \alpha_1\,E^2-\alpha_3\,|\vec
p\,|^2&=&\frac{m_D^2}3\eea from which \bea E_{A_L}&=&\frac{
1}{\sqrt 3}\sqrt{|\vec
p\,|^2\frac{\alpha_2}{\alpha_1}+\frac{m^2_D}{\alpha_1}}~\cr&&\cr
 E_{A_T}&=&\sqrt{|\vec
p\,|^2\frac{\alpha_3}{\alpha_1} +\frac{m^2_D}{3\alpha_1}}~, \eea
with
 \bea \alpha_2=\frac{\mu^2
g_s^2}{30 \Delta^2\pi^2}(a-9b)=-~\frac{\mu^2 g_s^2}{3240
\Delta^2\pi^2}\left(59-\frac{688}{3}\ln 2\right)~\cr
\alpha_3=-\frac{\mu^2 g_s^2}{30 \Delta^2\pi^2}(a+b)=-\frac{\mu^2
g_s^2}{3240\Delta^2\pi^2}\left(41-\frac{112}{3}\ln 2\right)~. \eea
We get easily the following results \be
m^R_{A_L}=m^R_{A_T}=m^R_{A_0}=\frac{m_D}{\sqrt{3\alpha_1}}\equiv
m_R ~.\ee Since $\alpha_3$ is large and negative, the inverse of
the penetration lenght looses its meaning for $A_T$. On the other
hand we get \be m^P_{A_L}=\frac{m_D}{\sqrt{\alpha_2}}=
\sqrt{\frac{3\alpha_1}{\alpha_2}}\,m^R \approx 3.73\,\Delta~.\ee
We get also \bea
m^*_{A_L}&=&\frac{\sqrt{3\alpha_1}}{\alpha_2}\,m_D~=~ 3\,
\frac{\alpha_1}{\alpha_2}m^R~\approx ~8.19\,\Delta\cr
m^*_{A_T}&=&\frac 1{\sqrt{3}}
\frac{\sqrt{\alpha_1}}{\alpha_3}\,m_D~=~
\frac{\alpha_1}{\alpha_3}\,m^R~ \approx~ -18.04\,\Delta~.\eea
Since  $m^*_{A_T}$ is negative, the spectrum of the
quasi-particles associated to the transverse gluons has a maximum
for $|\vec p|= 0$, which means that at very small temperatures,
which is the limit in which we work, these quasi-particles are
unlikely to be produced: this situation reminds the spectrum of
the  elementary excitations  of superfluid $He^4$ that, besides
the phonon and roton parts, presents a maximum for intermediate
momenta.

It can be useful to note that all these mass scales are determined
in terms of the common rest mass $m^R$ and of the ratios
$\alpha_1/\alpha_2\approx 1.61$ and $\alpha_1/\alpha_3\approx
-10.61$.

\section{In-medium fermions} To obtain the effective description
for the in-medium fermions, let us introduce for both the
left-handed and right-handed fermion fields
$\chi~(=\chi_L,~\chi_R)$  external fields
$\Psi~(=\Psi_L,~\Psi_R)$. By defining \be
\eta_\pm=P_{\pm}\Psi~,\ee and \be\Phi=\left(\matrix{\eta_+ \cr
C\eta^*_-}\right)~, \ee with $P_{\pm}$ given  in (\ref{iden}), we
can couple the external field $\Phi$ to the field $\chi$ (for
simplicity we consider here a single Fermi field). Then we have a
lagrangian density \be {\cal L} ={\cal L}_0+{\cal L}_{source}
=\int \frac {d\vec v_F}{8\pi} \left(\chi^\dagger
S^{-1}\chi+\chi^\dagger\Phi+\Phi^\dagger\chi\right)~,\ee with \be
S^{-1}=\frac 1 2\left[\matrix{ iV\cdot D & \Delta\cr\Delta
&i\tilde V\cdot D^*}\right]~. \ee Integrating over the field
$\chi$, or eliminating it through the equations of motion, we get
an effective action for $\Phi$ \be{\cal L}_{\rm eff}=\int \frac
{d\vec v_F}{8\pi}\frac 2{\Delta^2}
\Phi^\dagger\left[\matrix{i\tilde V\cdot D^* &-\Delta\cr -\Delta &
iV\cdot D}\right]\Phi~,\ee where we have made an expansion for low
momenta and small coupling constant. By re-expressing $\Phi$ in
terms of $\Psi$ and performing the integration over the Fermi
velocity we get
 \be{\cal L}_{\rm eff}=\left( \bar\Psi,~\bar
\Psi_c\right)\left[\matrix{ i\hat\gamma\cdot D^* & \Delta\cr\Delta
&i \hat \gamma\cdot D}\right]
 \left(\matrix{\Psi \cr \Psi_c}\right)
~,\ee
 where we have defined
\be \Psi_c=C\bar\Psi^T~, \ee and\be
\hat\gamma^\mu=(\gamma^0,~-~\frac{1}{3} \gamma^i)~,\ee and we have
scaled the field  as $\Psi$ \be \Psi\to\frac
{1}{\sqrt{2}\Delta}\Psi~.\ee One can notice that the origin of
the 1/3 in the lagrangian is again due to the average on the
Fermi velocity.

The dispersion relation for the field $\Psi$ is \be \tilde
p^2-\Delta^2=0,~~~\tilde p^\mu=(p^0,~\frac 1 3~\vec p)~,\ee
giving rise to \be p^0=\pm E,~~~E=\frac 1 3\sqrt{|\vec
p\,|^2+9\Delta^2}~.\ee We see that the rest mass is given by
$\Delta$. The velocity of in-medium fermions is \be \vec
v=\frac{\partial E}{\partial\vec p}=\frac 1 3\frac{\vec
p}{\sqrt{|\vec p\,|^2+9\Delta^2}}~.\ee and therefore, using the
same definition as in the previous Section for $m^*$ \be
m^*=\,9\,\Delta~.\ee It is tempting, even if not reliable within
our approximation, to consider the large $|\vec p\,|$ limit; for
$|\vec p\,|\gg 3\Delta$ one would get \be |\vec v\,|=\frac 1
3~,\ee to be compared to the velocity of the NGB's given by
$1/\sqrt{3}$.

\section{Conclusions}

After a presentation of the basic points of the formalism, we have
discussed the couplings of the goldstones and their effective
lagrangian. We have written down explicitly the needed trilinear
couplings of the goldstones to the fermions and the quadrilinear
couplings of two goldstones to two fermions. To explicitly obtain
the parameters of the effective action for the goldstones we have
performed a derivative expansion of the generating functional. In
this way we obtain the known results at one loop order. The value
of the velocity of the goldstones appears explicitly as following
from the integration over the Fermi velocity. We have then
discussed the  in-medium gluon polarization tensor. At the one
loop order the formalism reproduces the correct values for the
Debye and Meissner masses. The dispersion laws for both scalar and
vector gluons follow from the gluon field equations of motion in
the limit of high density; from them the various masses of the
scalar, longitudinal and transverse gluons have been computed,
i.e. the rest mass, the  effective masses and the inverse of the
penetration lenght. A main development has been the derivation of
the effective description for in-medium fermions. We have
introduced fermion external fields, both left- and right-handed,
to be coupled to the dynamical fermion fields. After integrating
over the dynamical fermions we have obtained an effective action
and the fermionic dispersion relation in terms of the gap
parameter. This provides for a value of the in-medium fermion
mass and shows again the role played by the integration over the
Fermi velocity.

The theory of color-flavor locking, as developed and studied by
many authors, has undoubtedly an intrinsic beauty and the high
density fermionic formalism used here provides for a simple and
appealing description. It remains to be seen to which physical
situations of matter under extreme conditions the theory can
usefully be applied, most probably by including additional
perturbations. \vskip.3cm\noindent {\bf{AKNOWLEDGMENTS}} We would
like to thank Krishna Rajagopal for an illuminating discussion
about a previous version of this paper.

\end{document}